\documentstyle{MYaipproc}
\begin{document}



\font\tenbit=cmmib10
\def\boldital{\fam 0 \tenbit}

\font\tenbb=msbm10
\def\bbten{\fam 0 \tenbb}

\font\sevenbb=msbm7
\def\bbsub{\fam 0 \sevenbb}

\font\teneu=eufm10
\def\euler{\fam 0 \teneu}


\def \ns{\enspace}
\def \ts{\thinspace}
\def \nts{\negthinspace}
\def \beq{\begin{equation}}
\def \eeq{\end{equation}}
\def \beqa{\begin{eqnarray}}
\def \eeqa{\end{eqnarray}}
\def \eps{\epsilon}
\def \vareps{\varepsilon}
\def \MeV{{\rm \enspace MeV}}
\def \GeV{{\rm \enspace GeV}}
\def \TeV{{\rm \enspace TeV}}
\def \trace{{\rm Tr}}
\def \half{\hbox{$1\over2$}}
\def \LABEL#1{\label{#1}}
 

\def \LQCD{\Lambda_{\rm QCD}}
\def \fLL{f_{\hbox{\bbsub LL}}}
\def \fLR{f_{\hbox{\bbsub LR}}}
\def \fRL{f_{\hbox{\bbsub RL}}}
\def \fRR{f_{\hbox{\bbsub RR}}}

\def \PLEFT{{\rm P}_{\hbox{\bbsub L}}}
\def \PRIGHT{{\rm P}_{\hbox{\bbsub R}}}


\title{Top Polarization at a $\mu^{+}\mu^{-}$ Collider\footnote{Talk 
presented by GM at the 5th International Conference on Physics 
Potential and Development of $\mu^{+}\mu^{-}$ Colliders, 
San Francisco, CA, December 15--17, 1999.}}
\author{Gregory Mahlon$^{*}$
and Stephen Parke$^{\dagger}$}
\address{$^{*}$Department of Physics, McGill University\\
3600 University St., Montr\'eal, QC H3A 2T8, Canada\\[0.25cm]
$^{\dagger}$Theoretical Physics Department, 
Fermi National Accelerator Laboratory\\
P.O. Box 500, Batavia, IL 60510, U.S.A.}


\maketitle

\begin{abstract}
The top quark pairs produced at a 
polarized muon 
collider are in a (nearly) pure spin configuration.
This result holds for all center-of-mass energies,
and is insensitive to the next-to-leading order 
QCD radiative corrections.
The decay products of a polarized top quark show strong
angular correlations.  We describe an interesting interference
effect between the left-handed and longitudinally polarized
$W$ bosons in top quark decay. 
This effect is easily observable in the angular distribution
of the charged lepton with respect to the beam axis.
\includegraphics{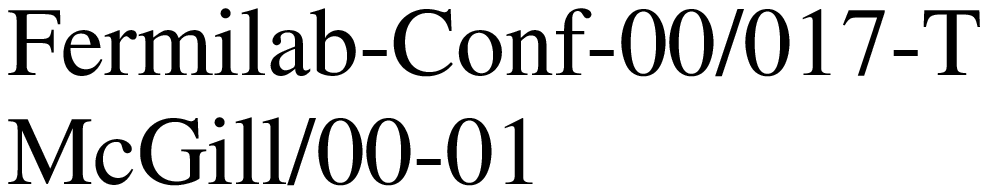}
\end{abstract}


With a mass of $173.8\pm5.2$ GeV\cite{PDG1999}, the top quark is 
by far the heaviest 
fermion in the Standard Model (SM).  Within the SM,  the top
quark decays very rapidly to a $W$ boson and a $b$ quark.
Since the $t$ decay width ($\Gamma_t \sim1.5$ GeV) is much greater than 
the spin decorrelation scale ($\Lambda_{\rm QCD}^2/m_t \sim 230$ keV),
a polarized $t$ quark decays before QCD can randomize its 
spin\cite{Bigi}.    
Thus, the decay products of polarized 
$t$ quarks exhibit strong angular correlations.

Because the focus of this talk is on a muon collider environment,
we should begin with a few words about the Higgs boson.
Although much has been said about the usefulness of a
muon collider as a ``Higgs factory'' for low-mass 
Higgs bosons, from the point of view of studying $t\bar{t}$
pair production, the Higgs boson does not play an important role.
A light SM Higgs boson shifts the total $t\bar{t}$ production
cross section at threshold by only a few percent\cite{TopThresh}.
This effect, which is comparable in size to the (incompletely known)
2-loop QCD corrections, becomes even smaller as the machine
energy is increased beyond the threshold region.
The other case to consider is a Higgs which is heavy
enough for the on-shell decay $h\rightarrow t\bar{t}$ to occur,
allowing for resonant $t\bar{t}$ production.  Assuming SM
couplings, a 400~GeV Higgs boson contributes
only a fraction of a percent to the total cross section through
this channel.
Consequently, as far as
top quark pair production is concerned, the same analysis applies
equally to $e^{+}e^{-}$ and $\mu^{+}\mu^{-}$ colliders.

The production of top quark pairs at a muon collider proceeds through
the $s$-channel via an off-shell photon or $Z$ boson.
For the purposes of our initial discussion, we will assume that
the $\mu^{-}$ beam is 100\% polarized in a left-handed helicity
state and that the $\mu^{+}$ beam is 100\% polarized in a right-handed
helicity state.  Details concerning the more general case with
arbitrary polarizations of the two beams may
be found in Refs.~\cite{ParkeShadmi,mumu97,DrellTalk}.
In the situation at hand, 
the matrix element for the
production of a $t\bar{t}$ pair whose spins are labelled by
$\lambda$ and $\bar\lambda$ takes the form
\beq
{\cal M} \sim {{e^2}\over{s}} \ts 
\Bigl[\bar{\hbox{\boldital v}}(\bar\mu) \ts 
\gamma^\mu \PLEFT \hbox{\boldital u}(\mu) \Bigr]
\Bigl[\bar{\hbox{\boldital u}}(t,\lambda)
\gamma^\mu \bigl\{ \fLL \PLEFT
+ \fLR \PRIGHT  \}
\hbox{\boldital v}(\bar{t},\bar\lambda) \Bigr],
\label{MatrixElement}
\eeq
where we have defined the chirality projection operators
$ \PLEFT \equiv \half(1{-}\gamma_5)$ and
$\PRIGHT \equiv \half(1{+}\gamma_5)$.
The effective couplings $\fLL$ and $\fLR$ incorporate the
different 
center-of-mass energy dependence of the photon and $Z$ boson 
propagators:
consequently they are weakly dependent on $\sqrt{s}$.
The ratio $\fLR/\fLL$ ranges
from 0.343 near threshold to 0.365 in the ultra-relativistic limit.
Note that since the muon is effectively massless at the center-of-mass
energies we are considering, the helicity of the muons is linked
to their chirality in the usual fashion.
On the other hand, both chiralities contribute
to the spin eigenstates
of the $t$ quark, even if we choose to work in the helicity
basis.  
Nevertheless, Parke and Shadmi~\cite{ParkeShadmi} have constructed
a spin basis, called the {\it off-diagonal basis}, in 
which more than 99\% of
the $t\bar{t}$ pairs are produced in the UD spin 
state at $\sqrt{s}=400 \GeV$.


\begin{figure}[b!] 
\includegraphics{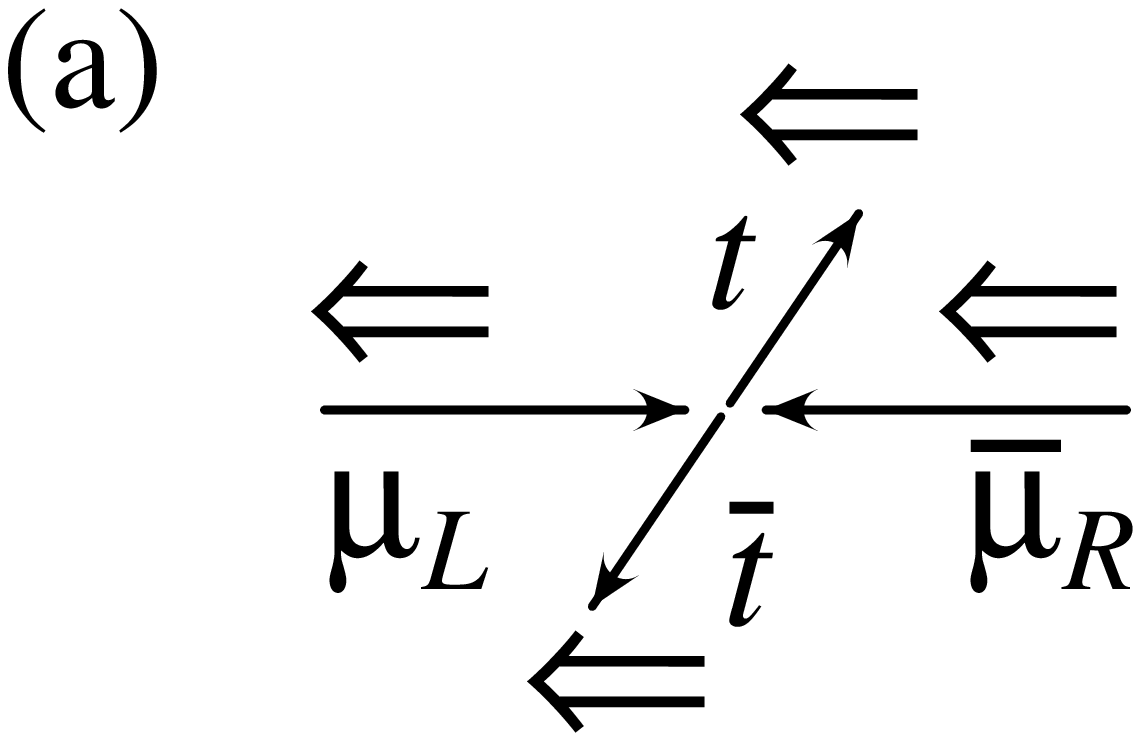}
\includegraphics{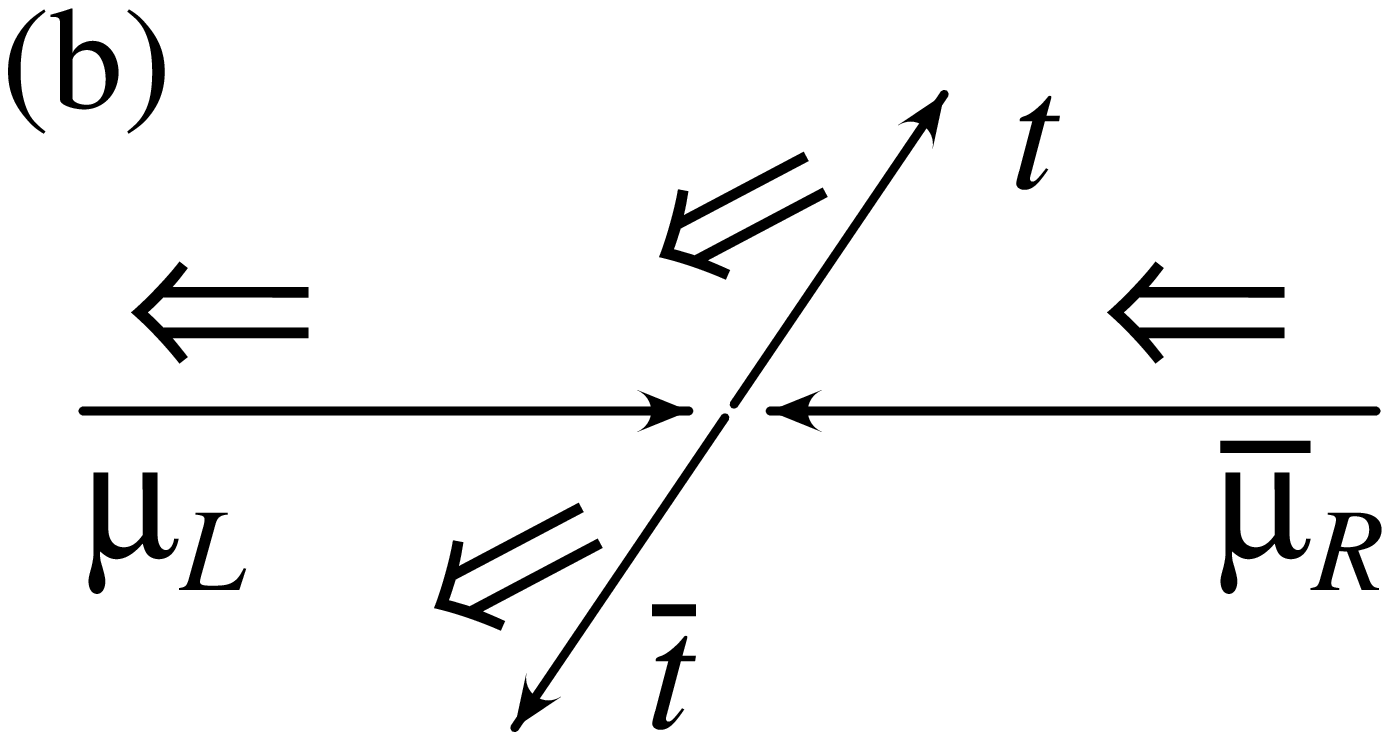}
\includegraphics{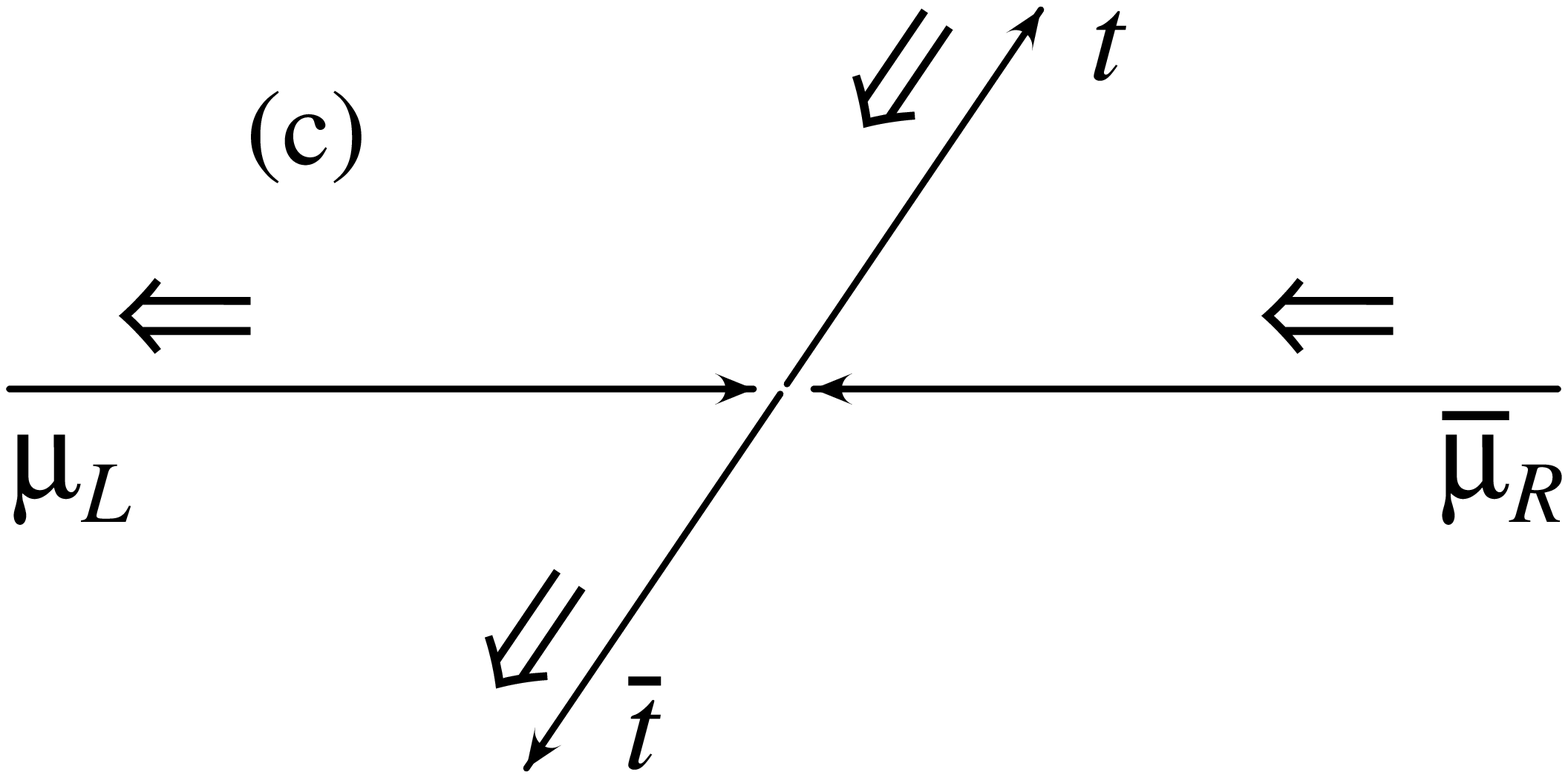}
\vspace{1.0in}
\caption{The process $\mu^{-}_L\mu^{+}_R\rightarrow t\bar{t}$
in the $\fLR\rightarrow0$ limit (a) near threshold, (b) at intermediate
energies, and (c) at ultra-relativistic energies.
}
\label{limits}
\end{figure}

In order to motivate the off-diagonal basis,
let us imagine that we can turn off the $\fLR$ coupling.  
Then, it is easy to see what happens in the two energy extremes
(Fig.~\ref{limits}).  
At ultrarelativistic energies, 
the left-handed chirality of the produced $t\bar{t}$ pair 
translates into left-handed helicity for
the $t$ and right-handed helicity for the $\bar{t}$:  we obtain
100\% $t_L \bar{t}_R$.  Near threshold, however, where
there is no orbital angular momentum involved, 
the $t$ and $\bar{t}$ spins must be parallel to the
beam axis,  since the initial
$\mu^{-}$ and $\mu^{+}$ spins add up to one unit along that direction.
At intermediate energies,
it is not surprising to learn that in the lab frame
the direction of the $t$ and $\bar{t}$ spins is not 
entirely along either the beam axis or the
$t\bar{t}$ production axis.  Instead, when
viewed in the $t$ rest frame, 
the $t$ spin is parallel to the $\mu^{+}$ momentum.
In the $\bar{t}$ rest frame, 
the $\bar{t}$ spin is antiparallel to the $\mu^{-}$ momentum.  
These associations ($t$ with $\mu^{+}$ and $\bar{t}$ with $\mu^{-}$)
may be understood by Fierzing the first term of 
Eq.~(\ref{MatrixElement}).
Thus, in this basis, the
$t\bar{t}$ spin state is 100\% UD
in the limit $\fLR\rightarrow0$. 

\begin{figure}[b!] 
\includegraphics{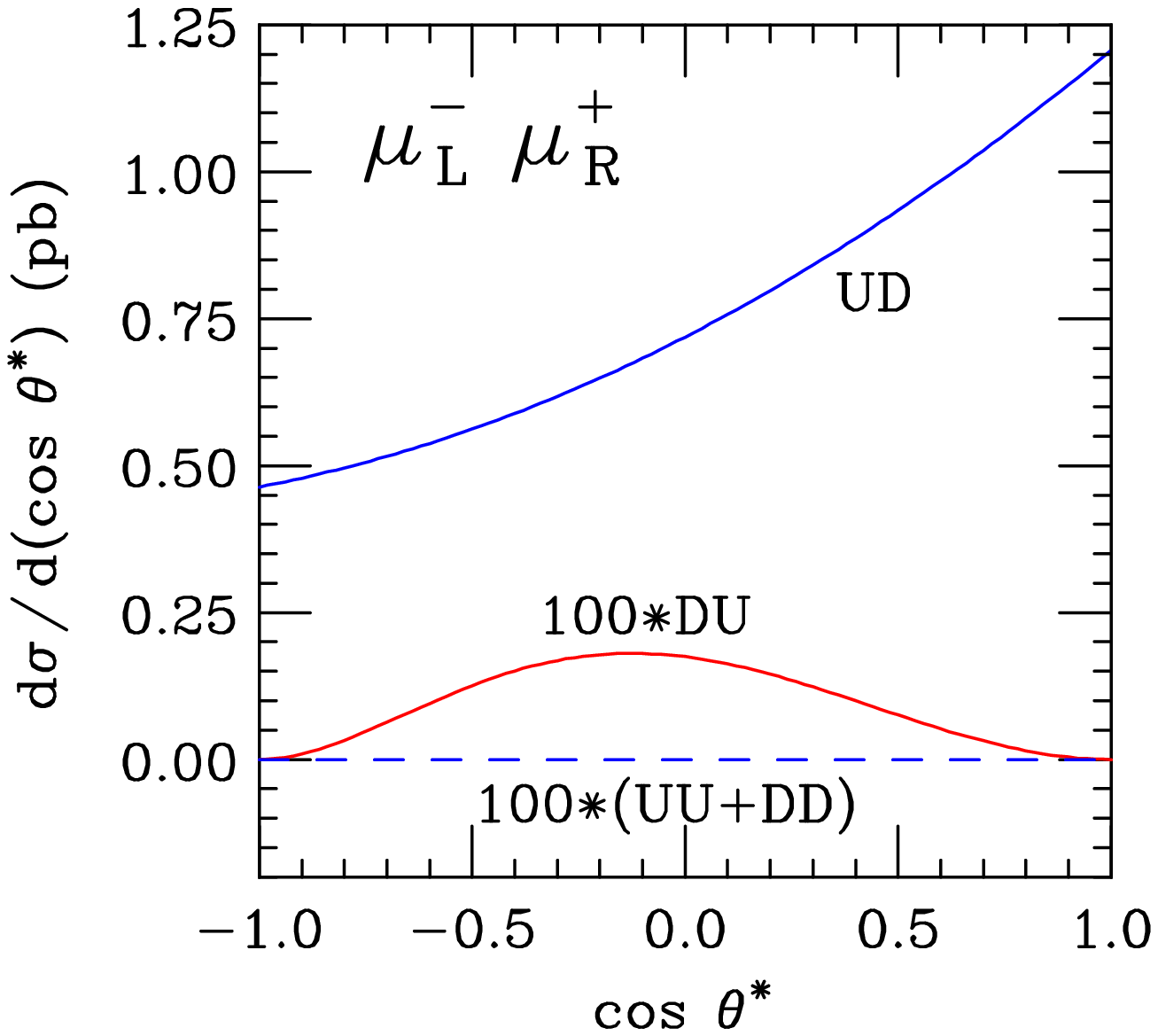}
\includegraphics{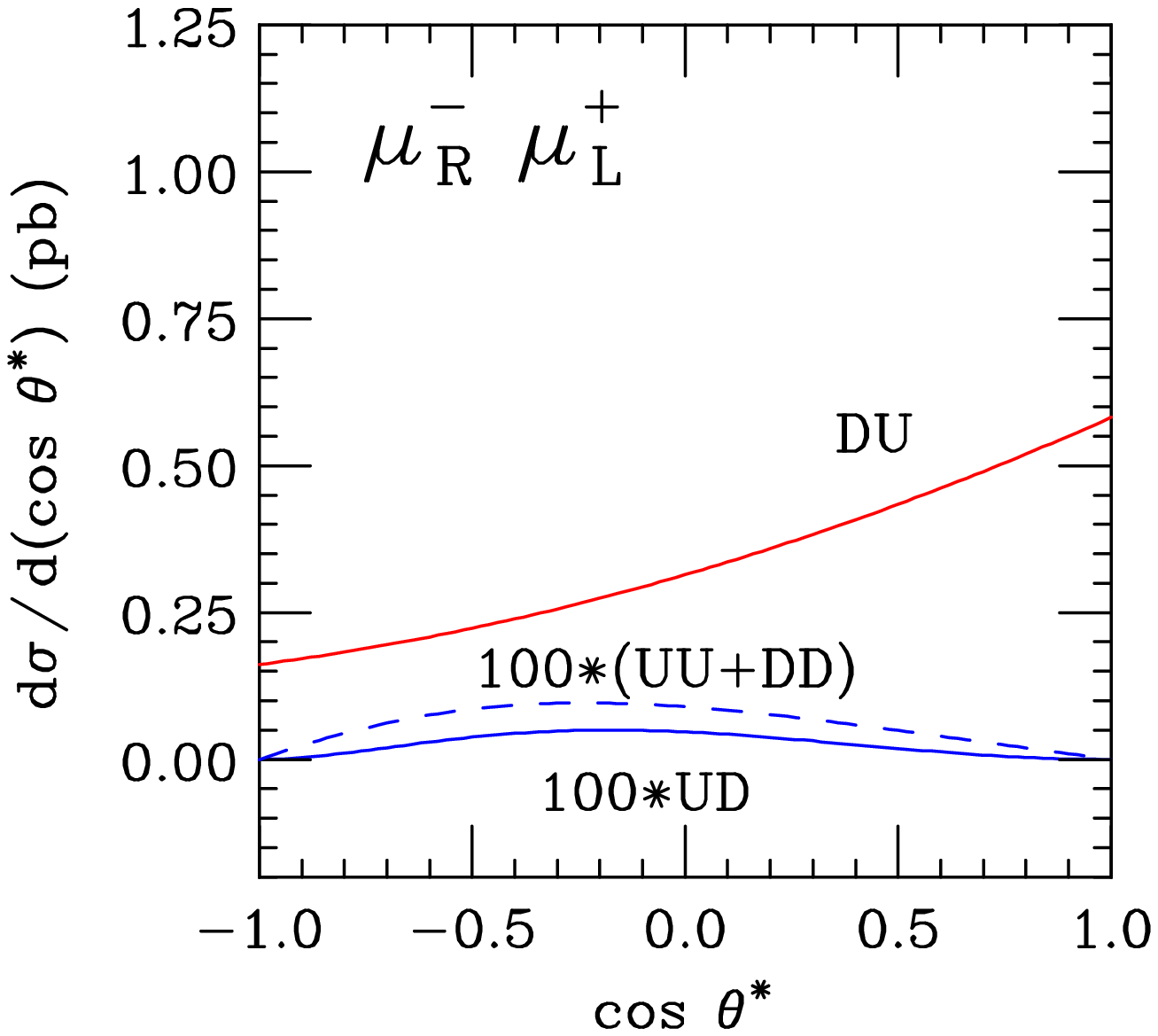}
\vspace{2.0in}
\caption{Top pair spin configurations 
as a function of the lab frame production angle
using the off-diagonal basis at $\protect\sqrt{s}=400\GeV$.
Note that the sub-dominant cross sections
have been multiplied by 100 to make them visible.  
}
\label{ODplot}
\end{figure}

Of course, we cannot set $\fLR=0$ in the real world.
However, because  
$\fLR/\fLL$ is not very large, it is possible to tweak the
spin axis choice a bit and retain the overwhelming dominance
of the UD spin configuration.  The result is the off-diagonal
basis of Parke and Shadmi\cite{ParkeShadmi}.  In the off-diagonal
basis, the UU+DD spin state vanishes identically for 
$\mu^{-}_L\mu^{+}_R$ collisions.  The DU spin state is suppressed
by $\beta^4(\fLR/\fLL)^2$.  Since the speed of the produced
top quarks is only $\beta\sim0.5$ at $\sqrt{s}= 400 \GeV$,
this is a significant suppression, and
$\mu_L^{-}\mu^{+}_R\rightarrow t_U\bar{t}_D$ accounts
for 99.88\% of the tree-level cross section.
Fig.~\ref{ODplot} shows the contributions to the $t\bar{t}$ cross
section as a function of the $t\bar{t}$ production angle in the lab
frame, decomposed into the various $t\bar{t}$ spin states for
a 400 GeV collider.


All of the preceding discussion has been at tree level.
Because the resulting spin state is so pure,
we might worry that the next-to-leading-order QCD corrections might
spoil this result, especially since the total (unpolarized)
cross section grows from 0.63 pb to 0.80 pb at $\sqrt{s}=400\GeV$,
an enhancement of more than 25\%.  
The spin dependence of the NLO QCD corrections has
been studied by Kodaira, Nasuno and Parke~\cite{OneLoopQCD},
who find that the off-diagonal basis defined at tree level 
is still the appropriate basis, and that the purity of the
$t_U\bar{t}_D$ state produced in $\mu_L^{-}\mu_R^{+}$ collisions
is still 99.85\%.  The physics behind this result is easily
understood.  The large enhancement in the total cross section
is mostly attributable to the production of soft gluons.
However, soft gluons are incapable of causing spin flips
of the top quark.
In the soft gluon limit,
matrix elements with an extra gluon factorize into 
matrix elements without the extra gluon times an eikonal
factor~\cite{SoftGluons}.  This factorization occurs
independently and with the {\it same}\ eikonal
factor for each spin configuration.
Thus, the spin composition of the produced $t\bar{t}$ pairs
is altered only slightly by the inclusion of NLO QCD effects.

\begin{figure}[b!] 
\includegraphics{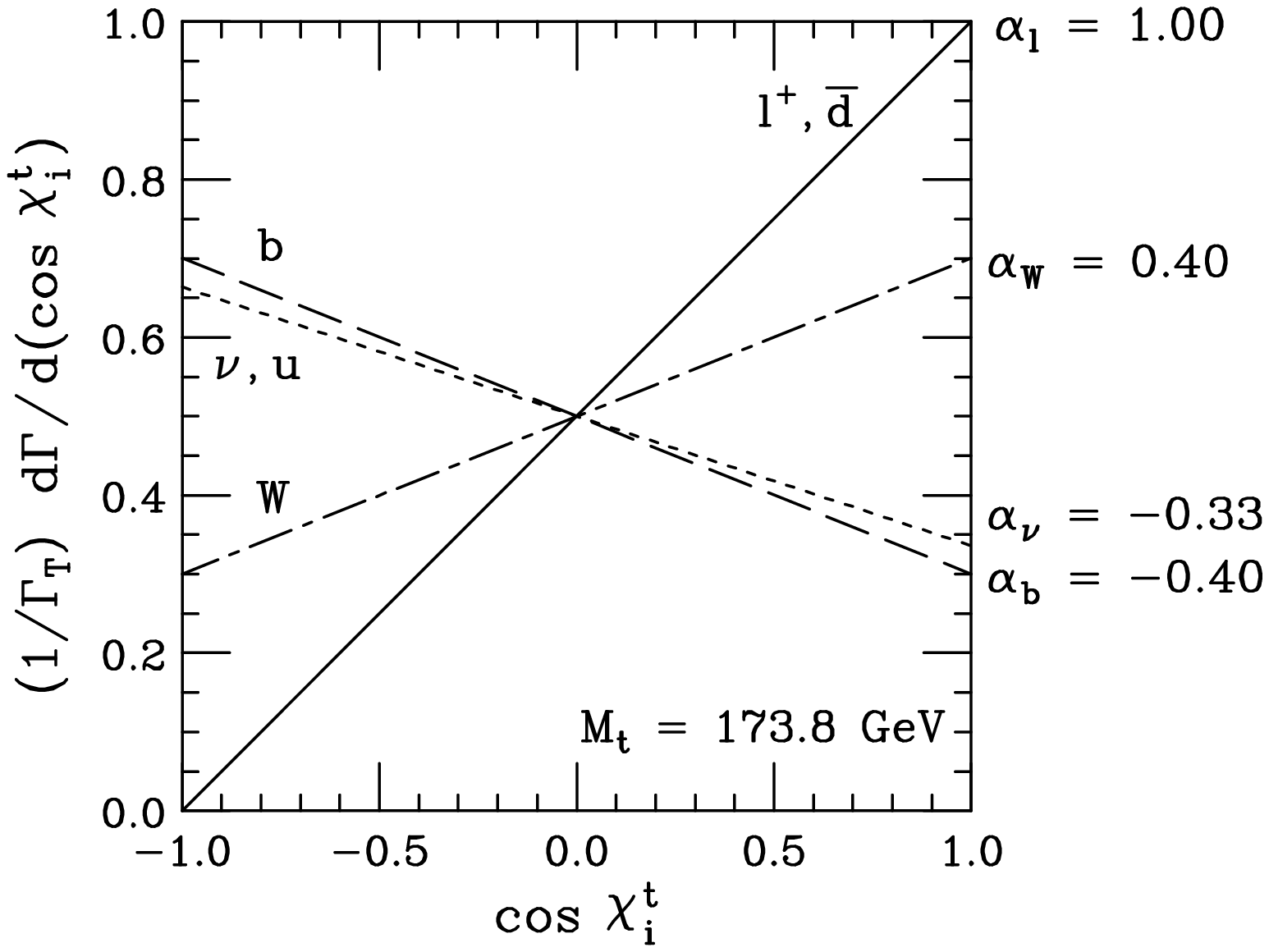}
\includegraphics{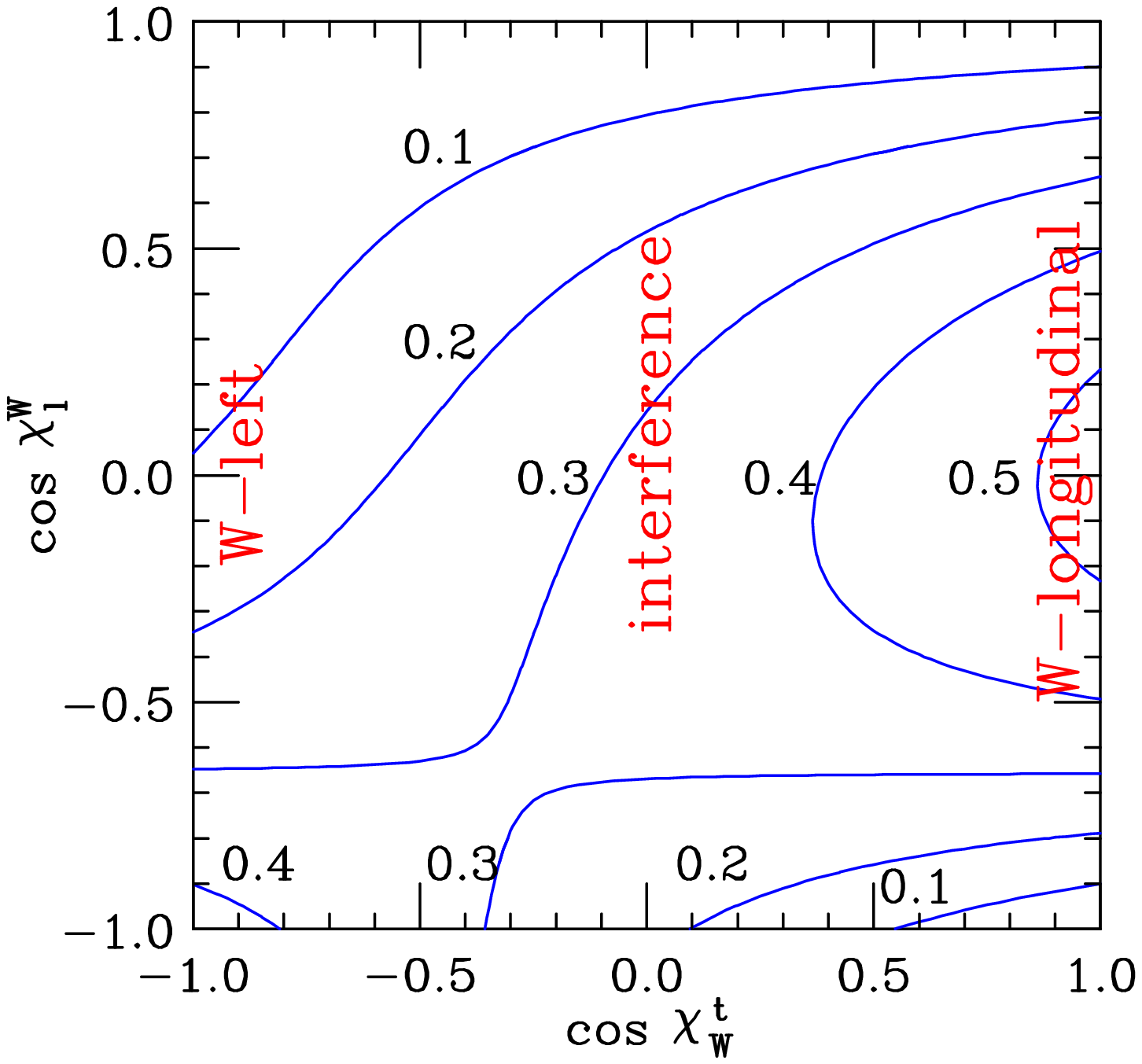}
\vspace{2.0in}
\begin{minipage}[b]{2.8in}
\caption{Angular correlations in the decay of a spin up top quark.
The lines labeled $W$, $b$, $\ell^{+}$, $\bar{d}$, $\nu$, and $u$
are the angle between the spin axis and the particle in the rest
frame of the top quark.
}
\label{ALPHA}
\end{minipage}
\hfill
\begin{minipage}[b]{2.8in}
\caption{Contours of the top quark decay distribution in
the $\cos\chi^t_W-\cos\chi_\ell^W$ plane.  $W$ bosons emitted
in the forward direction are primarily longitudinal, whereas
backward-emitted $W$'s are mostly left-handed.
}
\label{W2D}
\end{minipage}
\end{figure}

The decay products of a polarized top quark are strongly
correlated with the top quark spin axis.  These correlations
are most simply viewed in the $t$ rest frame, where  
the dependence upon the angle $\chi_i^t$ between the $i$th decay 
product and the $t$ spin axis is of the form
\beq
{{1}\over{\Gamma_T}}\ts
{ {d\Gamma} \over { d(\cos\chi_i^t) } }
= {{1}\over{2}}(1+\alpha_i\cos\chi_i^t)
\label{tdecay}
\eeq
for a spin up top quark\cite{DecayCorrel} (see Fig.~\ref{ALPHA}).
The charged
lepton or $d$-type quark from the decaying $W$ boson displays
maximal correlation with the top quark spin.  
The values of the $\alpha_i$'s for the other decay products
depend on the top and $W$ masses (see  
Ref.~\cite{xidep} for a convenient tabulation).

The spin states of the $W$ bosons coming from a spin up top 
quark decay  display interesting interference effects.  
The simplest description utilizes
the $W$ helicity measured in the $t$ rest frame.  Then, 
in the decay of a spin up top quark, 
the $W$'s are produced with left-handed or longitudinal helicity
only.  The ratio of left-handed to longitudinal $W$'s is 
$2m_W^2{:}m_t^2$ ($=30\%{:}70\%$), and is reflected in the decay 
angular distribution
\beq
{1 \over \Gamma_T}
{{d\Gamma}
\over
{d(\cos\chi_{\ell}^{W})}
}
= { 3\over4 } \thinspace
{ 
{\thinspace m_W^2 (1-\cos \chi_{\ell}^{W})^2 +
 m_t^2 \sin^2 \chi_{\ell}^{W}  \thinspace}
\over
{2m_W^2 + m_t^2} }.
\label{Wdecay}
\eeq
In Eq.~(\ref{Wdecay}), we have denoted 
$W$ rest frame angle 
between the $b$ quark and charged lepton by $\pi-\chi_\ell^W$.

In Fig.~\ref{W2D} we present a contour plot of the lepton emission
angle in the $W$ rest frame versus the $W$ emission angle in the
$t$ rest frame.  From this plot, we see that the longitudinal
$W$'s are emitted mainly parallel to the top quark spin,  while
the left-handed $W$'s are emitted mainly antiparallel to the top
quark spin.  Near $\cos\chi_W^t=0$, interference between the
left-handed and longitudinal $W$'s dominates.


\begin{figure}[t!] 
\includegraphics{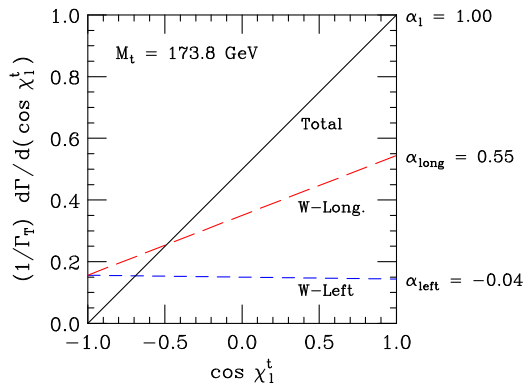}
\vspace{2.1in}
\caption{Interference effects between the left-handed and
longitudinal $W$ bosons from the decay of a spin up top quark.
Plotted is the angle between the charged lepton and the $t$
spin axis, broken down into decays through left-handed and
longitudinal $W$ bosons.  The solid line represents the quantum
mechanical sum of the two contributions.
}
\label{Left-Long-Inf}
\end{figure}

The interference between the left-handed and longitudinal $W$ bosons
from $t$ quark decay is the explanation for an apparent mystery
contained in Fig.~\ref{ALPHA}.  Recall that the $W$ bosons are
only moderately correlated with the top quark spin: $\alpha_W=0.40$.
However, one of the $W$ decay products (the charged lepton or
$d$-type quark) is maximally correlated with the top quark spin:
$\alpha_\ell = 1$!  In Fig.~\ref{Left-Long-Inf}, we show the
angular distribution of the charged leptons coming from left-handed
and longitudinal $W$ decays separately,\footnote{The $\alpha_i$'s
of Eq.~(\protect\ref{tdecay}) for the left-handed and longitudinal
$W$ bosons are given by
$$
\alpha_{\rm left} =
-{
{ \xi^3 + 8\xi^2 - 4\xi + 1 }
\over
{ (\xi-1)^3 }
}
+
{
{ 6\xi^3\ln\xi }
\over
{ (\xi-1)^4 }
} 
$$
and
$$
\alpha_{\rm long} =
{
{ (\xi+1)(\xi^2-8\xi+1) }
\over
{ (\xi-1)^3 }
} 
+
{
{ 12\xi^2 \ln\xi }
\over
{ (\xi-1)^4 }
}
$$
where $\xi \equiv (m_t/m_W)^2$.  For $m_t = 173.8 \GeV$ 
and $m_W=80.4 \GeV$ these expressions yield 
$\alpha_{\rm left} = -0.041$
and $\alpha_{\rm long} = 0.55$.
}
for the left-handed 
as well as their quantum
mechanical sum, including interference effects.  We see that at
$\cos\chi_\ell^t=-1$ there is total destructive interference.
For $\cos\chi_\ell^t>0$, the interference is, on average, constructive.
Thus, the interference term allows the charged lepton or $d$-type
quark to be more correlated with the top quark spin than the
parent $W$.


\begin{figure}[t!] 
\includegraphics{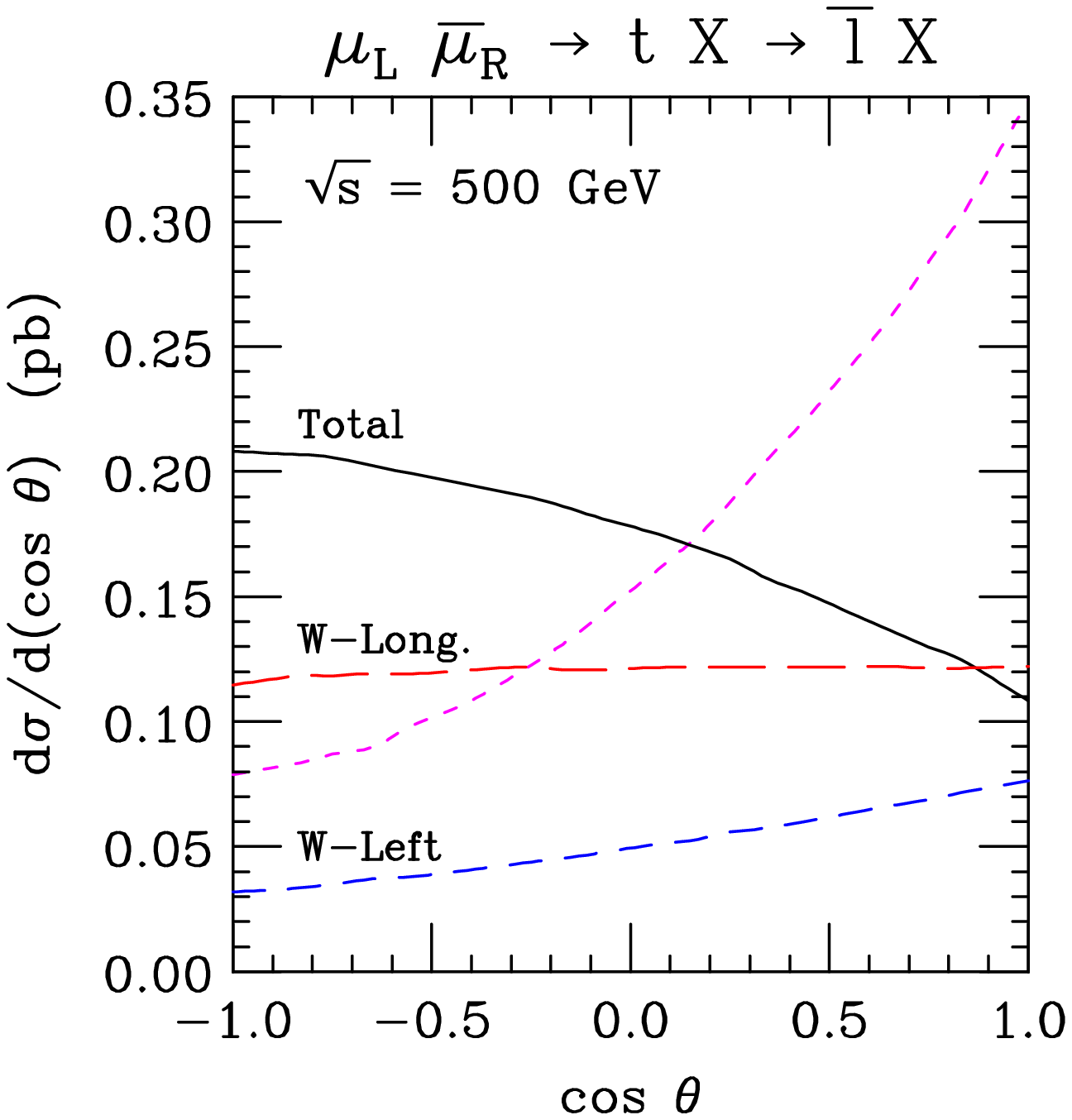}
\includegraphics{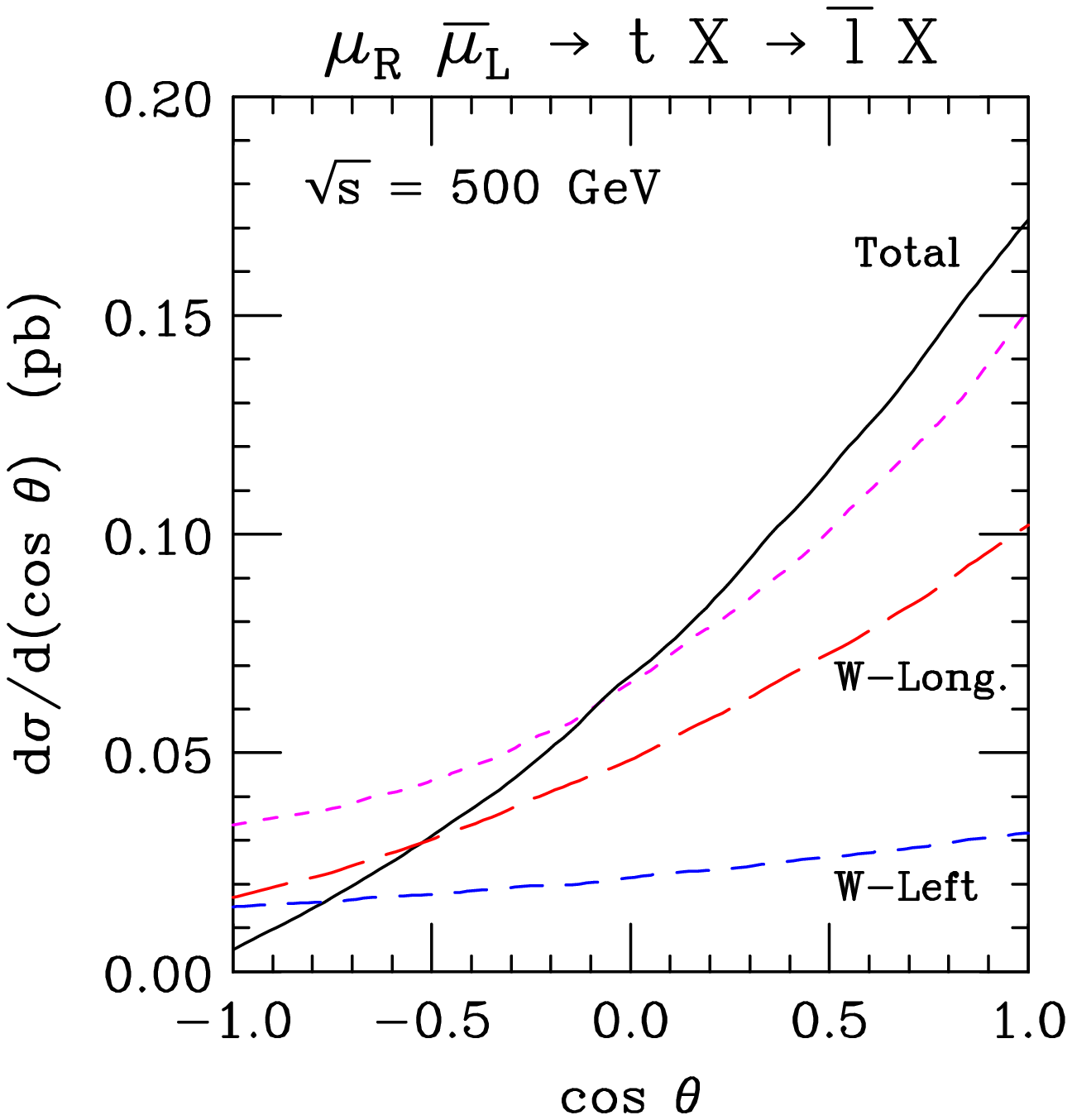}
\vspace{2.0in}
\caption{Lab frame charged lepton angular distributions which are
affected by the interference effects between the 
left-handed and longitudinal $W$ bosons.
Plotted is the angle between
the charged lepton and the $\mu^{-}$ beam direction in the lab
frame for the two $W$-boson helicity states (dashed lines) 
as well as their quantum-mechanical sum (solid line).
The dotted line indicates the top quark production
angular distribution. 
}
\label{LAB-Inf}
\end{figure}

This interference effect also leaves a visible imprint 
in the lab frame.  In Fig.~\ref{LAB-Inf} we show the distribution
of the angle between the charged lepton and the direction of
the $\mu^{-}$ beam, again broken down into the 
contributions from left-handed and longitudinal $W$ bosons.
The effect is particularly striking for
$\mu^{-}_L\mu^{+}_R$ collisions, as the shape of the quantum
mechanical sum of the two contributions is completely
dominated by presence of the interference term.  The dotted
line in Fig.~\ref{LAB-Inf} shows the top quark production angle.
Even though the $t$ quark is predominantly produced in the forward
direction in  
$\mu^{-}_L\mu^{+}_R$ collisions, 
the charged lepton from its decay tends to move backward in
the lab frame.
This is a consequence of the interference between the left-handed
and longitudinal $W$ bosons.


\end{document}